\documentstyle[eqsecnum,aps,preprint]{revtex}

\begin{document}
\preprint{Fermilab-pub-95/101-E}
\title{Measurement of the $WW\gamma$ gauge boson couplings
in $p\bar{p}$ Collisions at $\sqrt{s}=1.8$ TeV}
%
\author{
S.~Abachi,$^{12}$
B.~Abbott,$^{33}$
M.~Abolins,$^{23}$
B.S.~Acharya,$^{40}$
I.~Adam,$^{10}$
D.L.~Adams,$^{34}$
M.~Adams,$^{15}$
S.~Ahn,$^{12}$
H.~Aihara,$^{20}$
J.~Alitti,$^{36}$
% ...
\\
\vskip 0.50cm
\centerline{(D\O\ Collaboration)}
\vskip 0.50cm
}
\address{
\centerline{$^{1}$Universidad de los Andes, Bogot\'{a}, Colombia}
\centerline{$^{2}$University of Arizona, Tucson, Arizona 85721}
\centerline{$^{3}$Brookhaven National Laboratory, Upton, New York 11973}
\centerline{$^{4}$Brown University, Providence, Rhode Island 02912}
\centerline{$^{5}$University of California, Davis, California 95616}
\centerline{$^{6}$University of California, Irvine, California 92717}
\centerline{$^{7}$University of California, Riverside, California 92521}
\centerline{$^{8}$LAFEX, Centro Brasileiro de Pesquisas F{\'\i}sicas,
                  Rio de Janeiro, Brazil}
\centerline{$^{9}$CINVESTAV, Mexico City, Mexico}
\centerline{$^{10}$Columbia University, New York, New York 10027}
\centerline{$^{11}$Delhi University, Delhi, India 110007}
\centerline{$^{12}$Fermi National Accelerator Laboratory, Batavia,
                   Illinois 60510}
\centerline{$^{13}$Florida State University, Tallahassee, Florida 32306}
\centerline{$^{14}$University of Hawaii, Honolulu, Hawaii 96822}
\centerline{$^{15}$University of Illinois at Chicago, Chicago, Illinois 60607}
\centerline{$^{16}$Indiana University, Bloomington, Indiana 47405}
\centerline{$^{17}$Iowa State University, Ames, Iowa 50011}
\centerline{$^{18}$Korea University, Seoul, Korea}
\centerline{$^{19}$Kyungsung University, Pusan, Korea}
\centerline{$^{20}$Lawrence Berkeley Laboratory and University of California,
                   Berkeley, California 94720}
\centerline{$^{21}$University of Maryland, College Park, Maryland 20742}
\centerline{$^{22}$University of Michigan, Ann Arbor, Michigan 48109}
\centerline{$^{23}$Michigan State University, East Lansing, Michigan 48824}
\centerline{$^{24}$Moscow State University, Moscow, Russia}
\centerline{$^{25}$University of Nebraska, Lincoln, Nebraska 68588}
\centerline{$^{26}$New York University, New York, New York 10003}
\centerline{$^{27}$Northeastern University, Boston, Massachusetts 02115}
\centerline{$^{28}$Northern Illinois University, DeKalb, Illinois 60115}
\centerline{$^{29}$Northwestern University, Evanston, Illinois 60208}
\centerline{$^{30}$University of Notre Dame, Notre Dame, Indiana 46556}
\centerline{$^{31}$University of Panjab, Chandigarh 16-00-14, India}
\centerline{$^{32}$Institute for High Energy Physics, 142-284 Protvino, Russia}
\centerline{$^{33}$Purdue University, West Lafayette, Indiana 47907}
\centerline{$^{34}$Rice University, Houston, Texas 77251}
\centerline{$^{35}$University of Rochester, Rochester, New York 14627}
\centerline{$^{36}$CEA, DAPNIA/Service de Physique des Particules, CE-SACLAY,
                   France}
\centerline{$^{37}$Seoul National University, Seoul, Korea}
\centerline{$^{38}$State University of New York, Stony Brook, New York 11794}
\centerline{$^{39}$SSC Laboratory, Dallas, Texas 75237}
\centerline{$^{40}$Tata Institute of Fundamental Research,
                   Colaba, Bombay 400005, India}
\centerline{$^{41}$University of Texas, Arlington, Texas 76019}
\centerline{$^{42}$Texas A\&M University, College Station, Texas 77843}
}
\date{\today}
\maketitle
\begin{abstract}
The $WW\gamma$  gauge boson couplings were measured using
$p\bar{p}\rightarrow \ell\nu\gamma+X$  ($\ell=e,\mu$) events
at $\sqrt{s}=1.8$ TeV observed with the {D\O} detector at
the Fermilab Tevatron Collider.
The signal, obtained from the data corresponding to an integrated
luminosity of $13.8\ {\rm pb}^{-1}$, agrees well with
the Standard Model prediction.
A fit to the photon transverse energy spectrum
yields limits
at the $95\%$ confidence level on the CP--conserving anomalous coupling
parameters
of $-1.6<\Delta\kappa<1.8$ ($\lambda$ = 0) and
$-0.6<\lambda<0.6$ ($\Delta\kappa$ = 0).
\end{abstract}
Direct measurement of the $WW\gamma$  gauge boson couplings is possible
through study of  $W\gamma$ production in $p\bar{p}$ collisions at
$\sqrt{s}=1.8$ TeV.
The most general effective Lagrangian~\cite{Hagiwara},
invariant under $U(1)_{EM}$, for the $WW\gamma$ interaction
contains four coupling parameters, CP--conserving $\kappa$ and $\lambda$, and
CP--violating $\tilde{\kappa}$ and $\tilde{\lambda}$.
The CP--conserving parameters are related to the magnetic dipole ($\mu_W$) and
electric quadrupole ($Q_W^e$) moments of the $W$ boson, while
the CP--violating parameters are related to the electric dipole
($d_W$) and the magnetic quadrupole ($Q_W^m$) moments:
\mbox{$\mu_W=(e/2m_W)(1+\kappa+\lambda),$}
$Q_W^e=(-e/m_W^2)(\kappa-\lambda),\
d_W=(e/2m_W)(\tilde{\kappa}+\tilde{\lambda}),\
Q_W^m=(-e/m_W^2)(\tilde{\kappa}-\tilde{\lambda})$~\cite{Kim}.
In the Standard Model (SM) the $WW\gamma$ couplings at the tree level are
uniquely determined by
the $SU(2)_L\otimes U(1)_Y$ gauge symmetry:~$\kappa=1$
$(\Delta \kappa\equiv \kappa - 1 =0)$, $\lambda=0$,
$\tilde{\kappa}=0$, $\tilde{\lambda}=0.$
The direct and  precise measurement of the $WW\gamma$ couplings
is of interest since
the existence of anomalous couplings, i.e.~measured values different from the
SM predictions,  would indicate the presence of physics beyond the SM.
A $WW\gamma$ interaction Lagrangian with constant, anomalous couplings
violates unitarity at
high energies, and, therefore,
the coupling parameters must be modified to include
form factors (e.g.~$ \Delta\kappa (\hat{s})=
\Delta\kappa/(1+\hat{s}/\Lambda^2)^n,$
where $\hat{s}$ is the square of the invariant mass of the $W$ and the photon,
$\Lambda$ is the form factor scale, and
$n=2$ for a dipole form factor)~\cite{Baur}.

We present a measurement of the $WW\gamma$ couplings
using $p\bar{p}\rightarrow \ell\nu\gamma+X$ $(\ell=e,\mu)$ events
observed with the D\O\ detector\cite{NIM}
during the 1992--1993 run of the Fermilab Tevatron Collider,
corresponding to an integrated luminosity of
$13.8\pm 0.7\ {\rm pb}^{-1}.$
These events contain
the $W\gamma$ production process, $p\bar{p}\rightarrow W\gamma+X$ followed by
$W\rightarrow \ell\nu$, and
the radiative $W\rightarrow \ell\nu\gamma$ decay where the photon
originates from bremsstrahlung of the charged lepton.
Anomalous coupling parameters enhance the $W\gamma$ production
with a large $\hat{s}$, and thereby result in
an excess of events  with
high transverse energy, $E_T$, photons, well separated from the
charged lepton.
In the following, the electron and muon channels are referred to as
$W(e\nu)\gamma$ and $W(\mu\nu)\gamma$, respectively.

The D\O\ calorimeter system consists of uranium--liquid argon sampling
detectors
in  a central and two end cryostats, with a scintillator tile array in the
inter-cryostat regions.
The calorimeter~\cite{NIMTB} provides hermetic coverage
for $|\eta|< 4.4$ with  energy resolution of
$15\%/\sqrt{E({\rm GeV})}$ for electrons
and $50\%/\sqrt{E}$ for isolated pions, where $\eta$ is the pseudorapidity
defined as $\eta=-{\rm ln}(\tan(\theta/2))$, $\theta$ being the polar angle
with respect to the beam axis.
The calorimeter is read out in towers that subtend
$\Delta\eta\times\Delta\phi=0.1\times 0.1,$
$\phi$ being the azimuthal angle, and are
segmented longitudinally into 4 electromagnetic (EM)
and  4--5 hadronic  layers.
In the third EM layer, which typically contains $65\%$ of the EM shower energy,
the towers are subdivided transversely into
$\Delta\eta\times \Delta\phi=0.05\times 0.05.$
The central and forward drift chambers are used to identify charged tracks
for $|\eta|< 3.2.$
The muon system consists of magnetized iron toroids with one inner and two
outer layers of drift tubes, providing coverage for  $|\eta|< 3.3.$
The muon momentum resolution is
$\sigma(1/p)=0.18(p-2)/p^2\oplus 0.008$  with $p$ in ${\rm  GeV/c}$.

The $W(\ell\nu)\gamma$ candidates were obtained by
searching for events containing an isolated lepton
with high $E_T$,
large missing transverse energy, \mbox{$\not\!\!E_T$},
and an isolated photon.

The $W(e\nu)\gamma$ sample was
selected from events passing a trigger
which requires
an isolated EM cluster with $E_T>20$~GeV and
\mbox{$\not\!\!E_T>20$}~GeV.
This EM cluster was required to be within the fiducial region of the
calorimeter, $|\eta|< 1.1$ and at least 0.01 radians away from the azimuthal
boundaries between the 32 EM modules in the central calorimeter (CC),
or within $1.5< |\eta|< 2.5$ in the end calorimeters (ECs).
Kinematic selection  was made requiring
$E_T^e>25$~GeV, \mbox{$\not\!\!E_T> 25$}~GeV, and
$M_T>40\ {\rm GeV/c^2}$, where $M_T$ is the transverse mass
of the electron and the \mbox{$\not\!\!E_T$}  vector defined by
\mbox{$M_T=[2E_T^e\not\!\!E_T(1- \cos \phi^{e\nu})]^{1/2},$} and
$\phi^{e\nu}$ is the azimuthal angle
between the electron and the \mbox{$\not\!\!E_T$} vector.
The electron cluster must
(i) have a ratio of EM energy to the total shower energy
greater than 0.9;
(ii) have lateral and longitudinal shower shape consistent
with an electron shower~\cite{EID};
(iii) have the isolation variable of the cluster, $I$,
less than 0.15, where $I$ is defined as $I  = (E(0.4) - EM(0.2))/EM(0.2),$ and
$E(0.4)$ is the total calorimeter energy inside
a cone of radius
${\cal R}\equiv\sqrt{(\Delta\eta)^2+(\Delta\phi)^2}=0.4$, and
$EM(0.2)$ is the EM energy
inside a cone of 0.2 in the same units; and
(iv) have a matching track in the drift chambers.

The $W(\mu\nu)\gamma$ sample was selected from events passing
a trigger requiring an EM cluster with $E_T> 7$ GeV and a muon
track with transverse momentum, $p_T$, greater than 5~GeV/c.
A muon track was required to be in the region of $|\eta|< 1.7.$
Kinematic selection was made requiring  $p_T^{\mu}> 15$~GeV/c
and  \mbox{$\not\!\!E_T> 15$} GeV.
The muon track must
(i) have hits in the inner drift-tube layer;
(ii) have a good track fit in the muon system;
(iii) traverse a minimum field integral of $2.0\ {\rm Tm};$
(iv) have a time within $100\ \rm{ns}$ of the beam crossing;
(v) have an impact parameter, computed using only hits in the muon system,
smaller than 22 (15) cm in the bend  (non-bend) view;
(vi)  be isolated from a nearby jet in $\eta$--$\phi$ space,
$\Delta  R_{\mu-{\rm jet}}>0.5;$ and
(vii) have a matching track in the
drift chambers as well as a minimum energy deposition of 1 GeV in the
calorimeter.
To reduce the background due to $Z\gamma$ production,
events were rejected if they contained an extra muon track with
$p_T^\mu> 8$~GeV.

The requirements on photons  were common to both the
$W(e\nu)\gamma$ and the $W(\mu\nu)\gamma$ samples.
We required $E_T^\gamma > 10$ GeV and
the same geometrical and quality selection as  for electrons,
except that we required a tighter isolation, $I< 0.10$, and
that there be no track matching the calorimeter cluster.
In addition, we required that the separation between a photon
and a lepton be ${\cal R}_{\ell\gamma}> 0.7.$
This requirement suppresses the contribution of the radiative $W$ decay
process, and
minimizes the probability for a photon cluster to merge with a nearby
calorimeter cluster associated with an electron or a muon.
The above selection criteria yielded 11 $W(e\nu)\gamma$ candidates
and 12 $W(\mu\nu)\gamma$~candidates.

The background estimate, summarized in Table~\ref{table1},
includes contributions from:
$W+{\rm jets}$, where a jet is misidentified as a photon;
$Z\gamma$, where the $Z$ decays to
$\ell^+\ell^-$, and one of the leptons is undetected or is
mismeasured by the detector and contributes to \mbox{$\not\!\!E_T;$}
$W\gamma$ with $W\rightarrow \tau\nu$ followed by
$\tau\rightarrow \ell\nu\bar{\nu}.$
We estimated the $W+{\rm jets}$ background
using
the  probability, ${\cal P}(j\rightarrow ``\gamma ")$,
for a jet to be misidentified as a photon determined
as a function of $E_T$ of the jet
by measuring the fraction of jets in a sample of multijet events
that pass our photon identification requirements.
The contribution from direct photon events in the multijet sample to
${\cal P}(j\rightarrow ``\gamma ")$ was subtracted using
a conversion method~\cite{DG}.
We found the
misidentification probability to be ${\cal P}(j\rightarrow ``\gamma ")
\sim 4\times 10^{-4}$
($\sim 6\times 10^{-4}$)  in the CC (EC)
in the $E_T$ region between  10 and 40 GeV,
where our photon candidates occur.
By applying ${\cal P}(j\rightarrow ``\gamma ")$
to the observed $E_T$ spectrum of jets in the
inclusive $W(\ell\nu)$ sample,
we calculated the total number of $W+{\rm jets}$ background
events to be $1.7\pm 0.9$ and $1.3\pm 0.7$
for $W(e\nu)\gamma$ and
$W(\mu\nu)\gamma$, respectively, where the uncertainty is dominated by
the uncertainty in ${\cal P}(j\rightarrow ``\gamma ")$
due to the direct photon subtraction.
We tested for a bias in the $W+{\rm jets}$ background estimate
due to a possible difference in jet fragmentation (e.g.~the number of
$\pi^0$s in a jet)
between jets in the $W$ sample
and those in the multijet sample
by parameterizing
${\cal P}(j\rightarrow ``\gamma ")$  further as a function of
the EM energy fraction of the jet and found no evidence for a bias.
Because the $W+{\rm jets}$ background is computed using
observed $W(\ell\nu)$ events, it includes
the background originating from  $\ell$+jets,
where $\ell$ is a jet misidentified as an electron,
a cosmic ray muon or a fake muon track.

The backgrounds due to $Z\gamma$ and $W\rightarrow \tau\nu$
were estimated using
the $Z\gamma$ event generator  of Baur and Berger~\cite{Berger} and
the {\small ISAJET} program~\cite{ISAJET}, respectively,
followed by a full detector simulation using the {\small GEANT}
program~\cite{GEANT}.
Subtracting the estimated backgrounds
from the observed number of events,
we found the number of signal events to be
$$N_{\rm sig}^{W(e\nu)\gamma}=9.0^{+4.2}_{-3.1}\pm 0.9,
{}~~~N_{\rm sig}^{W(\mu\nu)\gamma}
=7.6^{+4.4}_{-3.2}\pm 1.1,$$
where the first uncertainty is statistical, calculated following
the prescription for Poisson processes with background given in
Ref.~\cite{PDG},
and the second is systematic.

The trigger and offline lepton selection efficiencies,
shown in Table~\ref{table2}, were estimated using
$Z\rightarrow \ell\bar{\ell}$ and  $W\rightarrow \ell\nu$ events.
The detection efficiency for photons with $E_T>25$ GeV was determined
using electrons from $Z$ decays.
For photons with lower $E_T$ there is a decrease in detection efficiency
due to  the cluster shape requirement, which was determined
using test beam electrons, as well as
the isolation requirement, which was
determined by measuring the energy in a cone of radius
${\cal R}=0.4$ randomly placed in the inclusive $W(e\nu)$ sample.
Combining this $E_T$--dependent efficiency with the probabilities
of losing a photon due to $e^+e^-$ pair conversions,
$0.10~(0.26)$ in the CC~(EC),  and due to
an overlap with a random track in the event,
$0.065~(0.155),$
we estimated that the overall photon selection efficiency is
$0.43\pm 0.04$ ($0.38\pm 0.03$) at $E_T^\gamma=10$ GeV, and that it increases
to $0.74\pm 0.07$ ($0.58\pm 0.05$) for $E_T^\gamma> 25$ GeV.

\begin{figure}
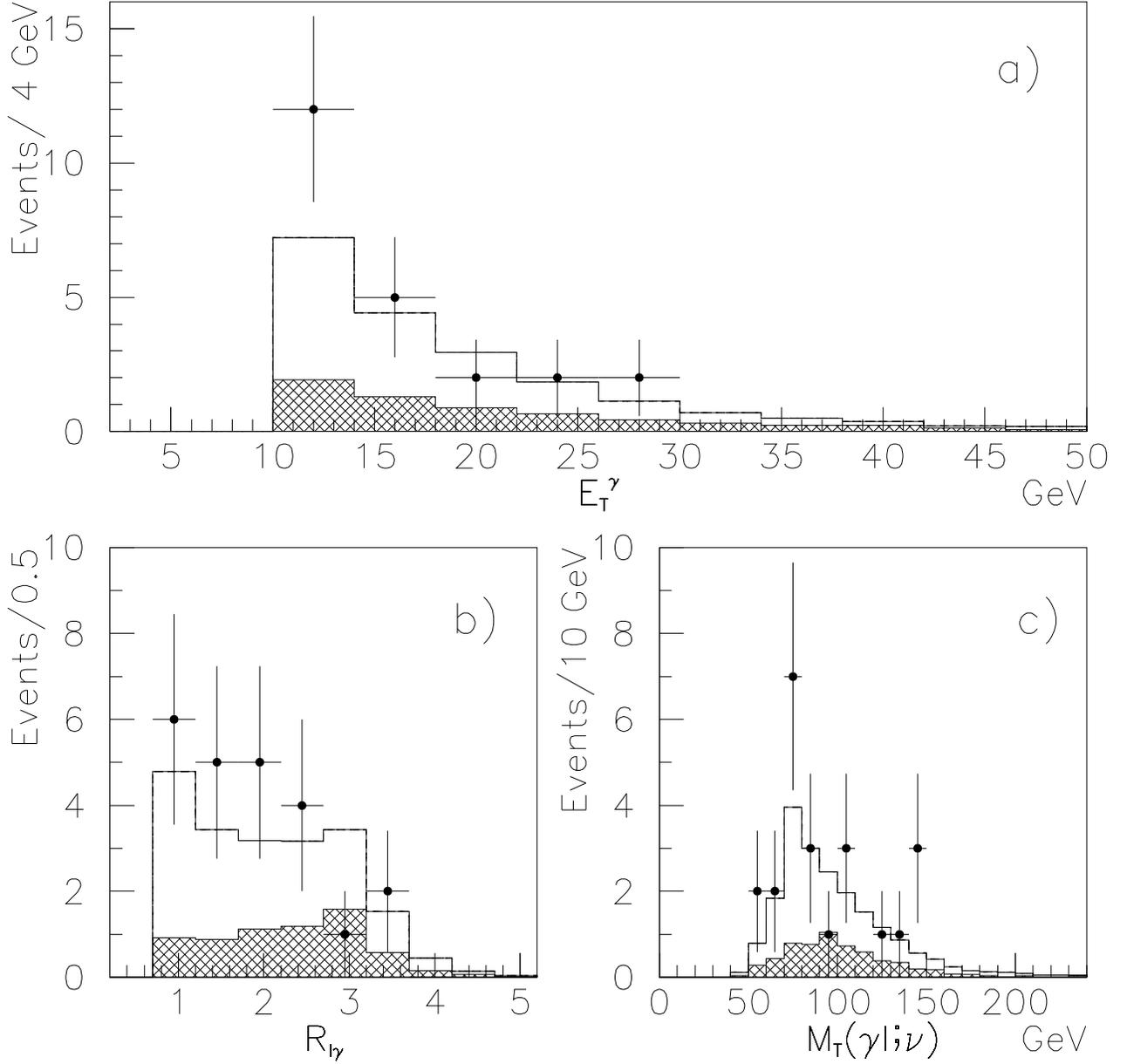

\caption{Distribution of (a)~$E_T^\gamma$,
(b)~${\cal R}_{\ell\gamma}$ and (c)~$M_T(\gamma\ell;\nu)$
for the $W(e\nu)\gamma$ + $W(\mu\nu)\gamma$
combined sample.
The points are data. The shaded areas represent the estimated
background, and the solid histograms are the expected signal from the Standard
Model  plus the estimated background.}
\label{ET}
\end{figure}

We calculated the kinematic and geometrical acceptance
as a function of coupling parameters
using the Monte Carlo program
of Baur and Zeppenfeld~\cite{Baur2}, in which the $W\gamma$ production
and radiative decay processes are generated to leading order, and
higher order QCD effects are approximated by a K-factor of 1.335.
We used the ${\rm MRSD\_^\prime}$ structure functions~\cite{MRSD}  and
simulated the $p_T$ distribution of the $W\gamma$ system using the observed
$p_T$ spectrum of the $W$ in the inclusive $W(e\nu)$ sample.
Using the acceptance for  SM couplings of  $0.11\pm 0.01$ for $W(e\nu)\gamma$
and $0.29\pm 0.02$ for $W(\mu\nu)\gamma$ and the efficiencies quoted above,
we calculated the $W\gamma$ cross section (for photons with $E_T^\gamma> 10$
GeV
and ${\cal R}_{\ell\gamma}>0.7$) from a combined $e+\mu$ sample:
$\sigma(W\gamma)=138^{+51}_{-38}(stat)\pm 21(syst)\ \ {\rm pb},$
where the systematic uncertainty includes
the uncertainty ($11\%$) in the $e/\mu/\gamma$
efficiencies,
the uncertainty ($9.1\%$) in the choice of the structure functions,
the $Q^2$ scale at which the structure functions are evaluated and
the $p_T$ distribution of the $W\gamma$ system,
and the uncertainty ($5.4\%$) in the integrated luminosity
calculation.
The observed cross section agrees with the SM prediction
of $\sigma_{W\gamma}^{SM}=112\pm 10\ {\rm pb}$ within errors.
Figure~\ref{ET} shows
the data and the SM prediction plus the background
in the distributions of $E_T^\gamma,$  ${\cal R}_\ell\gamma,$
and
the cluster transverse mass
defined by
$M_T(\gamma\ell;\nu)=(((m_{\gamma\ell}^2+|{\bf E_T^\gamma}
+{\bf E_T^\ell}|^2)^{\frac{1}{2}}+{\not\!\!E_T})^2-|{\bf E_T^\gamma}+
{\bf E_T^\ell}+{\bf \not\!\!E_T}|^2)^{\frac{1}{2}}.$
Of the 23 events we observed,
11 events having $M_T(\gamma\ell;\nu)\leq M_W$  are primarily the
radiative $W$ decay events plus background.

\begin{figure}
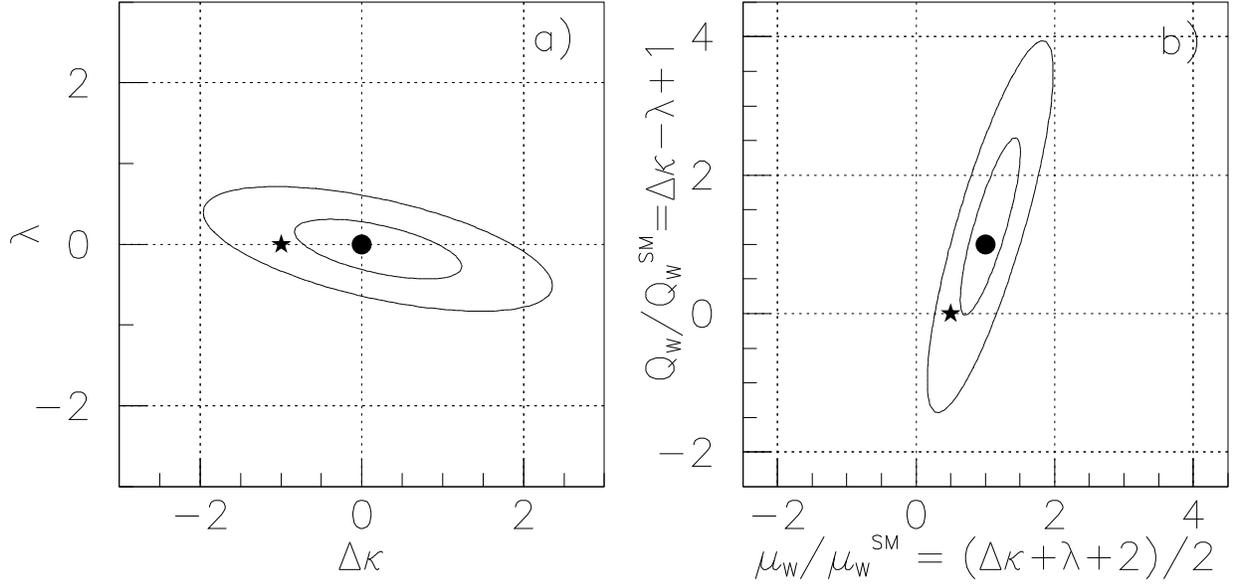

\caption{Limits on (a) CP--conserving anomalous coupling
parameters $\Delta \kappa$ and $\lambda,$ and on (b) the magnetic dipole,
$\mu_W$, and electric quadrupole, $Q^e_W$, moments.
The ellipses represent the $68\%$ and $95\%$ CL exclusion contours.
The symbol,~$\bullet$, represents the Standard Model values, while
the symbol,~$\star$, indicates
the $U(1)_{EM}$--only coupling of the $W$ boson to a photon,
$\Delta\kappa=-1$ and $\lambda=0$ ($\mu_W=e/2m_W$ and $Q^e_W=0$).}
\label{CONT}
\end{figure}

To set limits on the anomalous coupling parameters,
a binned maximum likelihood fit was performed
on the  $E_T^\gamma$ spectrum for each of the $W(e\nu)\gamma$ and
$W(\mu\nu)\gamma$ samples, by calculating the
probability for the sum of the Monte Carlo prediction and the background
to fluctuate to the observed number of events.
The uncertainties in background estimate, efficiencies, acceptance
and integrated luminosity
were convoluted  in the likelihood function with Gaussian distributions.
A dipole form factor with a form factor scale $\Lambda=1.5$ TeV
was used in the Monte Carlo event generation.
The limit contours for the CP--conserving
anomalous coupling parameters $\Delta \kappa$ and
$\lambda$ are shown in Fig.~\ref{CONT}, assuming that the CP--violating
anomalous coupling parameters $\tilde{\kappa}$ and
$\tilde{\lambda}$ are zero.
We obtained limits at the $95\%$ confidence level (CL) of
$$-1.6<\Delta\kappa<1.8\ (\lambda=0),~~~-0.6<\lambda<0.6\ (\Delta\kappa=0)$$
for $\hat{s}=0$ (i.e.~the static limit).
The $U(1)_{EM}$--only coupling of the $W$ boson to
a photon, which leads to  $\kappa=0\ (\Delta\kappa=-1)$ and $\lambda=0$, and
thereby, $\mu_W=e/2m_W$ and $Q_W^e=0\cite{TDLEE},$
is excluded at the $80\%$ CL, while
the zero magnetic moment ($\mu_W=0$) is excluded at the $95\%$ CL.
Similarly, limits on CP--violating coupling parameters  were obtained
as $-1.7<\tilde{\kappa}<1.7\ (\tilde{\lambda}=0)$ and
$-0.6<\tilde{\lambda}<0.6\ (\tilde{\kappa}=0)$ at the $95\%$ CL.
We studied the form factor scale dependence of the results
and found that the limits are insensitive to the form factor for
$\Lambda>200$~GeV and
are well within the constraints imposed by the S-matrix
unitarity~\cite{Unitarity} for $\Lambda=1.5$~TeV.
We also performed a two dimensional fit including ${\cal R}_\ell\gamma$, and
found that the results are within $3\%$ of those obtained from a fit to
the  $E_T^\gamma$ spectrum only.
Our results represent the currently most stringent
direct limits on anomalous $WW\gamma$ couplings~\cite{UA2},~\cite{CDF}.

We thank U.~Baur and D.~Zeppenfeld for providing
us with the $W\gamma$ Monte Carlo
program and for many helpful discussions.
We thank the Fermilab Accelerator, Computing, and Research Divisions, and
the support staffs at the collaborating institutions for their contributions
to the success of this work.
We also acknowledge the support of the
U.S. Department of Energy,
the U.S. National Science Foundation,
the Commissariat \`a L'Energie Atomique in France,
the Ministry for Atomic Energy and the Ministry of Science and
Technology Policy in Russia,
CNPq in Brazil,
the Departments of Atomic Energy and Science and Education in India,
Colciencias in Colombia, CONACyT in Mexico,
the Ministry of Education, Research Foundation and KOSEF in Korea
and the A.P. Sloan Foundation.

\newpage
\begin{table}
\caption{Summary of
$W(e\nu)\gamma$ and $W(\mu\nu)\gamma$ data
and backgrounds.}
\label{table1}
\begin{tabular}{lcc}
 & $W(e\nu)\gamma$ & $W(\mu\nu)\gamma$ \\
\tableline
Source: & & \\
$~~~W+{\rm jets}$ & $1.7\pm 0.9$   & $1.3\pm 0.7$ \\
$~~~Z\gamma$      & $0.11\pm 0.02$ & $2.7\pm 0.8$ \\
$~~~W(\tau\nu)\gamma$ & $0.17\pm 0.02$ & $0.4\pm 0.1$ \\
Total Background        & $2.0\pm 0.9$ & $4.4\pm 1.1$ \\
\tableline
Data & 11 & 12
\end{tabular}
\end{table}
\begin{table}
\caption{Summary of trigger ($\epsilon_{trig}$) and
lepton selection ($\epsilon_\ell$) efficiencies.}
\label{table2}
\begin{tabular}{lcccc}
 &\multicolumn{2}{c}{$W(e\nu)\gamma$} & \multicolumn{2}{c}{$W(\mu\nu)\gamma$}
 \\
\tableline
  & & & & \\
  & $|\eta|< 1.1$ & $1.5< |\eta|< 2.5$ & $|\eta|< 1.0$ &
 $1.0<|\eta|< 1.7$ \\
\tableline
  & & & & \\
$\epsilon_{trig}$  &
$0.98\pm 0.02$ & $0.98\pm 0.02$  & $0.74\pm 0.06$ & $0.35\pm
0.14$ \\
$\epsilon_\ell$ & $0.79\pm 0.02$ & $0.78\pm 0.03$ &
$0.54\pm 0.04$ & $0.22\pm 0.07$ \\
\end{tabular}
\end{table}


\end{document}